\documentclass{kluwer}
\usepackage{graphicx} 
\begin{document}
\begin{article}
\begin{opening}
\title{Similarities and differences between coronal holes and the quiet
Sun: are loop statistics the key?}
\author{T. \surname{Wiegelmann}\email{wiegelmann@linmpi.mpg.de}}
\author{S.K. \surname{Solanki}} 
\institute{Max-Planck-Institut f\"ur Sonnensystemforschung
\footnote{Former Max-Planck-Institut f\"ur Aeronomie},
Max-Planck-Strasse 2, 37191 Katlenburg-Lindau, Germany}

\date{DOI: 10.1007/s11207-004-3747-2 \\
Bibliographic Code: 2004SoPh..225..227W }


\runningtitle{Coronal Holes}
\runningauthor{Wiegelmann and Solanki}

\begin{ao}
Kluwer Prepress Department\\
P.O. Box 990\\
3300 AZ Dordrecht\\
The Netherlands
\end{ao}

\begin{motto}

\end{motto}
\begin{abstract}
Coronal holes (CH) emit significantly less at coronal
temperatures than quiet Sun regions (QS),
but can hardly be distinguished in
most chromospheric and lower transition region lines.
A key quantity for the understanding of this phenomenon is
the magnetic field. We use data from SOHO/MDI to reconstruct
the magnetic field in coronal holes and the quiet Sun
with the help of a potential magnetic model. Starting from
a regular grid on the solar surface we then trace field
lines, which provide the overall geometry of the 3D magnetic
field structure. We distinguish between open and closed
field lines, with the closed field lines being assumed
to represent magnetic loops. We then
try to compute some properties of coronal loops.
The loops in the CH are found to be on average flatter than in the QS.
High and long closed loops are extremely rare, whereas short and low-lying
loops are almost as abundant in coronal holes as in the quiet Sun.
When interpreted in the light of loop scaling laws this result suggests
an explanation for the relatively strong chromospheric and transition
region emission (many low-lying, short loops), but the weak coronal emission
(few high and long loops) in coronal holes. In spite of this contrast
our calculations also
suggest that a significant fraction of the cool emission in CHs comes from
the open flux regions. Despite these insights provided by the magnetic field
line statistics further work is needed to obtain a definite answer to
the question if loop statistics explain the differences between coronal holes
and the quiet Sun.
\end{abstract}


\abbreviations{\abbrev{KAP}{Kluwer Academic Publishers};
   \abbrev{compuscript}{Electronically submitted article}}

\nomenclature{\nomen{KAP}{Kluwer Academic Publishers};
   \nomen{compuscript}{Electronically submitted article}}

\classification{JEL codes}{D24, L60, 047}
\end{opening}
\section{Introduction}
\label{sec1}
Coronal holes \cite{waldmeier57,waldmeier75},
 are regions with a significantly
reduced emissivity in lines corresponding to coronal temperatures.
Usually coronal holes are divided into two classes.
During solar activity minimum large polar
coronal holes exist with opposite magnetic polarity in the northern and
southern hemisphere.
Around solar activity maximum smaller isolated coronal holes are
present at different latitudes, even close to the equator.
However, the main property of coronal holes, the reduction in
brightness of coronal radiation is common to both types.
For the rest of the paper we distinguish between  quiet Sun regions without
coronal holes (and call them simply {\it Quiet Sun regions QS})
and quiet Sun regions inside coronal holes (which we call {\it Coronal holes CH}).

An intriguing observation is that the emission from chromospheric and transition
region lines formed at temperatures below $6 \; 10^5 K$
is not significantly reduced in coronal holes
\cite{wilhelm00b,stucki00,stucki02,xia03} with the exception of He lines, whose
formation is strongly affected by coronal radiation.
Here we use a simple statistical analysis of the magnetic
structure to see if a cause for this behaviour can be found, viz. that coronal
holes can easily be identified in radiation alone at $10^6 K$, but are not
seen below $6 \; 10^5 K$.

\begin{figure}
\hspace*{\fill}
\includegraphics[clip,height=7cm,width=12cm]{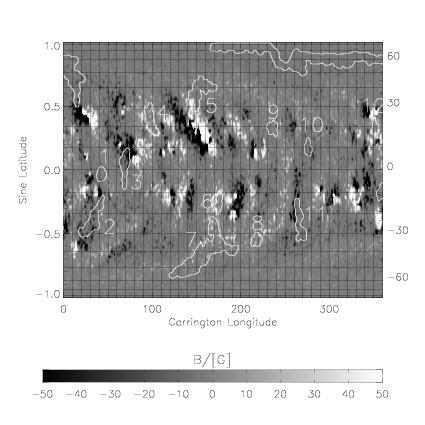}
\caption{A synoptic chart (09.04.2001-06.05.2001) of the magnetic
field constructed from SOHO/MDI. The boundaries of
coronal holes taken from He I measurements made at Kitt Peak are overlaid.
The analysed coronal holes are numbered.}
\label{fig1}
\end{figure}

\begin{figure}
\hspace*{\fill}
\includegraphics[clip,width=11cm]{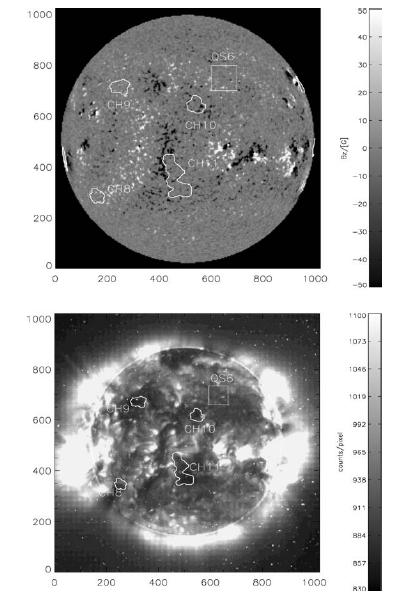}
\caption{Full disk MDI magnetogram (top panel) and EIT Fe XII,
$19.5 nm$ image (bottom panel) for
16.4.2001 at 02:30h,
with boundaries of coronal holes overlaid.
Also marked is an analysed quiet Sun region.
The disc center corresponds to a Carrington longitude of $272^{\circ}$.}
\label{fig2}
\end{figure}

Coronal holes are also visible in absorption lines.
The $1083$ absorption line in He I is
very useful for the identification of coronal holes, because
it has a greatly reduced absorption there
\cite{harvey75} and this line has the additional advantage that
it can be observed from the ground. The coronal maps used in this paper
(see Figs. \ref{fig1} and \ref{fig2})
have been created at NSO/Kitt Peak from these He I observations
\cite{kharvey01}.

The key quantity for the understanding of coronal holes is
the magnetic field. According to current understanding coronal holes
differ from the normal quiet Sun mainly through the structure of the
magnetic field, with the field lines being mainly closed in the normal
quiet Sun and open above coronal holes
(e.g. \citeauthor{altschuler72} \citeyear{altschuler72}).
Hot gas is trapped in closed loops, but is able to escape along the
open field lines. The trapped gas radiates, causing the normal quiet
Sun to be brighter.

Although this basic difference in magnetic structure can qualitatively
explain the decreased coronal brightness in coronal holes, it begs the
question why transition region brightness is not different as well. One
possibility is that the chromospheric and transition region are located
near the footpoints of the magnetic field lines and their properties are
independent of the properties of the overlying corona. One argument against this
idea come from the fact that in such a model the transition region is heated by
downward conduction of heat from the corona. If there is significantly less
coronal gas, as in a coronal hole it is surprising that the transition region
is not affected.
Another problem with such models was pointed out by, e.g.,
\inlinecite{dowdy86}.
The emission measure predicted by models such as that of \cite{gabriel76}
falls far short of the observations at lower temperatures.
\inlinecite{dowdy86} proposed that short, cool loops are common
in the quiet Sun. They harbour
a significant fraction of the gas at lower transition region temperatures.
Evidence for such loops has been found by \citeauthor{feldman98}
(\citeyear{feldman98,feldman01}).

Since the temperature of a loop depends on its length
$(T \propto L^{1/3})$;
\citeauthor{rosner78} (\citeyear{rosner78}), \citeauthor{kano96}
(\citeyear{kano96})
short loops are expected to be cool,
while long loops are expected to be hot.
Thus, one way of explaining the decreased coronal emission together with the
unaffected transition region emission in coronal holes is by having a
greatly decreased number of long loops, but an almost equal number of short
loops relative to the normal quiet Sun. Here we test this hypothesis.

The magnetic field strength in the photosphere beneath coronal holes
has been measured by several
authors, (e.g.,
\citeauthor{levine77} \citeyear{levine77},
\citeauthor{bohlin78} \citeyear{bohlin78},
\citeauthor{harvey82} \citeyear{harvey82},
\citeauthor{deforest97} \citeyear{deforest97},
\citeauthor{belenko01} \citeyear{belenko01}).
The magnetic flux in coronal holes clearly shows a dominant polarity.
This is consistent with coronal holes being the
source region of the fast solar wind flowing on open magnetic field lines.
\inlinecite{harvey82} found an average magnetic field strength
in coronal holes of 3 to 36 G near
the solar activity maximum and 1 to 7 G close to the minimum.
\citeauthor{zhang02} (\citeyear{zhang02,zhang03}) found a range from about 4 to
30 G, where polar coronal holes are more close
to the lower and solar maximum coronal holes more close to the higher limit
of this range.
The magnetic field is, however, not exclusively unipolar in coronal holes and
consequently coronal
holes should contain also locally closed coronal loops besides the open flux
\cite{levine77}.

The presence of both magnetic polarities  inside
the coronal hole  leads to the formation of closed magnetic loops.
Here we compare features of
these closed loops in coronal holes with quiet Sun loops by extrapolating
the magnetic field starting from photospheric magnetograms.
We organize the paper as follows. In section \ref{sec2} we describe
the method for computing the magnetic field from photospheric magnetograms.
In section \ref{sec3} we compare the average height and length of
magnetic loops for several coronal holes and quiet Sun regions. The
distribution functions of loop length and heights are investigated in
section \ref{sec4}, while in section  \ref{sec5} we compute loop temperatures
with the help of scaling laws. In section \ref{sec6} we discuss which magnetic
features are responsible for coronal holes and finally  we
draw conclusions and give an outlook for future work in section \ref{sec7}.
\section{Reconstruction of the magnetic field  in the solar atmosphere}
\label{sec2}
\begin{figure}
\hspace*{\fill}
\includegraphics[clip,width=14cm]{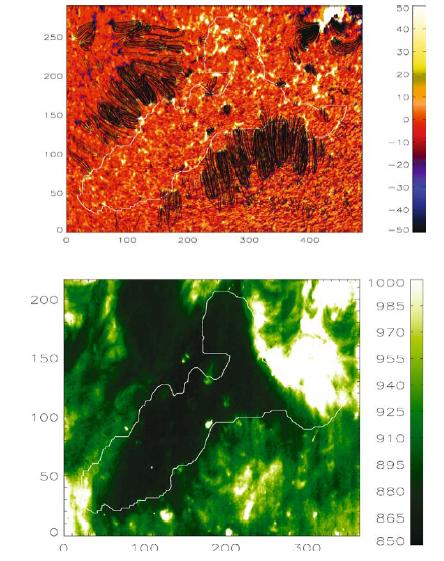}
\caption{Top panel: MDI magnetogram (colour) with overlaid
(black) closed magnetic loops in a coronal holes and the
neighbouring quiet Sun.
We only plot field lines with a magnetic
field strength $B_{\rm lim} \ge 20G$ in order to allow individual
loops to be distinguished. In the actual calculations a lower threshold
of $2.7G$ is used.
The coronal hole is outlined in white.
Bottom panel: EIT $19.5 nm$ image of the same region. Both images show
the number of pixels on the x and y axis. Please note the different pixel
size of MDI ($1 pixel=1.978 arc sec$) and EIT ($1 pixel=2.629 arc sec$).}
\label{fig3}
\end{figure}
Direct measurements of the magnetic field
in the chromosphere and corona are mainly restricted to large active
regions (e.g.
\citeauthor{kundu01} \citeyear{kundu01},
\citeauthor{white02} \citeyear{white02},
\citeauthor{sami03} \citeyear{sami03},
\citeauthor{lagg04} \citeyear{lagg04}),
although a few
measurements of the magnetic field in the quiet corona are available
(e.g. \citeauthor{lin98} \citeyear{lin98}, \citeauthor{raouafi02} \citeyear{raouafi02}).
This kind of data is only available for a few individual cases and usually one
has to calculate the coronal magnetic field by means of an extrapolation
from photospheric field measurements. The extrapolation method requires some
assumption regarding electric currents in the solar atmosphere.
The simplest approach is to compute current free or potential magnetic fields
(e.g. \citeauthor{schmidt64} \citeyear{schmidt64},
\citeauthor{semel67} \citeyear{semel67}).
Potential fields can be reconstructed from the line of sight photospheric
magnetic field alone. The corresponding photospheric magnetic field
measurements are available e.g. from the SOHO/MDI line-of-sight magnetograph
\cite{scherrer95}.
 Force free and non-force-free magnetic field
models are mathematically more difficult and need additional observational
data, e.g. photospheric vector magnetic fields, optical observation of
coronal plasma structures or even, additionally, the tomographicly
reconstructed density structure for non-force-free configurations.
For  low lying loops in the quiet Sun
investigated in this work we
concentrate on potential fields. Coronal currents have a significant
effect on the structure of long (say, some $100 Mm$ long) loops in active
regions, but for the very short (some $10 Mm$ or less) quiet Sun loops investigated
here their influence can be neglected.

We use a Greens function method to compute the potential magnetic field
in the solar atmosphere
(see \citeauthor{aly89} \citeyear{aly89}, and appendix \ref{appendixb} for details.)
We investigated $12$ coronal hole and $8$ quiet Sun regions during Carrington
rotation 1975 (09.4.-06.05.2001).
This period was employed because 56 consecutive 1-min magnetograms are
available every day, from which 56 min averages are constructed, after
compensating for solar (differential) rotation
(see \citeauthor{krivova04} \citeyear{krivova04}).
 The magnetic field data were taken from
full disc MDI magnetograms when the coronal hole was closest to the central
meridian (See table \ref{tableA} in appendix \ref{appendixb}).
As an example, we show the potential magnetic field lines
for coronal hole region 7 and the surrounding area in figure \ref{fig3}, top panel.
The image displays only the closed loops with a lower limit of the magnetic
field strength $20 G$.
Open field lines are not shown.
Outside the coronal hole (thick white line) the loop density is significantly
higher and the loops are also longer than in the hole. The bottom panel of
figure \ref{fig3} shows the same region in the EIT Fe XII $19.5 nm$ line.
In the coronal hole both the emissivity in EIT as well as the number and length
of coronal loops are significantly reduced, although in the rightmost and
bottom left edges extended bright structures are found within the CH
boundary determined from 1083 He I observations.

A crucial step is how to draw conclusions about the properties
of coronal loops from magnetic field lines. Due to the high conductivity
the coronal plasma is frozen into the magnetic field. Consequently the
emitting plasma structures also outline the magnetic field lines. The
gradients of e.g. the plasma pressure and temperature perpendicular to
the magnetic field are much higher than along the magnetic field.
Within the statistical study it is assumed that these properties are
constant along a given field line and only changes from one bundle
of field lines to the next.  This assumption seems to be
not so bad and is consistent with CDS observations regarding the thermal
structure of hot coronal loops \cite{brkovic02}. Theoretically the number
of magnetic field lines in the corona (and even the number of
field lines in one observed plasma loop)  is infinite.
To undertake statistical studies regarding the length and height
of coronal loops we make the following assumptions. Each magnetic
field line starts at the center of a MDI pixel. We assume that
this field line is representative for all magnetic field lines
starting within the area of the corresponding MDI pixel on the
photosphere. Each such representative closed field line is
assumed to also represent a magnetic loop. Consequently the
number of magnetic loops is determined by the fraction
of MDI-pixels that correspond to closed field lines.
Now, it can be argued that this is an arbitrary measure of a
loop. This is true, but given that from an extrapolation of
the magnetic field alone no conclusions can be drawn regarding
the distribution of the coronal plasma transverse to the field
lines, this is a reasonable assumption. An MDI pixel is the
smallest spatial scale which is accessible to us for the
extrapolation and the assumption that each pixel supports
a separate loop is driven by the small-scale structure seen
by TRACE and the NIXT instrument (e.g. \citeauthor{golub90}
\citeyear{golub90}).
Although individual plasma loops may well be formed from
field lines gathered from different sized areas on the solar
surface, our approach should in a statistic sense provide
the correct distribution of loop lengths, heights, etc.
A bias could be introduced into the comparison between
quiet Sun and coronal hole only if on average loops in one
of these regions are thicker than in the other, for which we
have found no evidence in the literature.
\section{Analysis of loop height and length}
\label{sec3}
\begin{table}
\caption{Table of coronal holes (CH) and quiet Sun regions (QS).
The first column identifies the region, the second column shows the spatially
averaged
net magnetic field $\langle B_z  \rangle$ in $G$, the third column the relative flux balance
$\frac{\langle B_z  \rangle}{\langle |Bz| \rangle}$, the fourth column the area
of the analysed feature,
the fifth column the average height $H$ (in Mm) of closed loops,
the sixth column the average length $L$ of closed loops (in Mm), the second last
column the number $n$ of closed loops and the last column the density of loops.
CH1* and QS3* are not included in the further analysis. Unsigned values have been
used when forming the average (excluding CH1 and QS3).}
\begin{tabular}{|r|r|r|r|r|r|r|r|}
\hline
& $\frac{\langle B_z \rangle}{[G]}$ & $\frac{\langle B_z \rangle}{\langle |B_z| \rangle}$&
$\frac{Area}{Mm^2} $& $\frac{H}{[Mm]}$ & $\frac{L}{[Mm]}$ & $n$ & $\frac{n}{Area}$\\
\hline
CH1* &$-0.8$&$-0.25$&$7497$&$3.7$&$20.8$& 911 &$0.12 $\\
CH2 &$-7.9$&$-0.82$&$25071$&$0.7$&$7.0$&2009 &$0.08 $\\
CH3 &$-14.3$&$-0.91$&$16863$&$1.7$&$17.6$&949 &$0.06$\\
CH4 &$8.3$&$0.90$&$14435$&$0.5$&$4.0$&607 &$0.04$\\
CH5 &$2.7$&$0.62$&$19471$&$0.9$&$7.1$&1594 &$0.08$\\
CH6 &$17.2$&$0.95$&$7998$&$0.7$&$10.8$&310 &$0.04$\\
CH7 &$3.7$&$0.58$&$65077$&$0.8$&$5.2$&9199 &$0.14$\\
CH8 &$8.8$&$0.83$&$7117$&$0.5$&$3.7$& 574 &$0.08$\\
CH9 &$2.2$&$0.59$&$7189$&$1.6$&$9.7$&552 &$0.08$\\
CH10&$-6.8$&$-0.86$&$6049$&$0.8$&$6.4$&271 &$0.04$ \\
CH11&$-3.0$&$-0.59$&$17895$&$0.8$&$6.6$&1866 &$0.10$ \\
CH12&$8.8$&$0.78$&$4349$&$0.6$&$7.6$&253 &$0.06$ \\
$\langle \mbox{CH} \rangle$ & $7.6$&$0.77$&$17410$&$0.9$&$7.8$&$1650$&$0.07$  \\
&$$&$$&$$&$$&$$& &$$ \\
QS1 &$-0.1$&$-0.02$&$43904$&$3.0$&$15.1$&10122 &$0.23$\\
QS2 &$ 0.5$&$0.10$&$15876$&$4.3$&$22.0$&2653 &$0.17$\\
QS3* &$-4.3$&$-0.62$&$37632$&$0.9$&$6.4$&5815 &$0.15$\\
QS4 &$-0.1$&$-0.02$&$7056$&$3.1$&$13.5$&914 &$0.13$\\
QS5 &$1.2$&$0.29$&$7056$&$2.5$&$11.4$&1129 &$0.16$\\
QS6 &$0.3$&$0.09$&$7056$&$3.3$&$15.2$&1254 &$0.18$\\
QS7 &$0.4$&$0.07$&$11760$&$2.3$&$9.2$&3580 &$0.30$\\
QS8 &$-0.1$&$-0.03$&$7056$&$2.8$&$12.6$&1382&$0.20$\\
$\langle \mbox{QS} \rangle$&$0.4$&$0.09$&$14252$&$3.0$&$14.1$&$3005$&0.20 \\
\hline
\end{tabular}
\label{table2}
\end{table}
\begin{figure}
\hspace*{\fill}
\includegraphics[width=14cm]{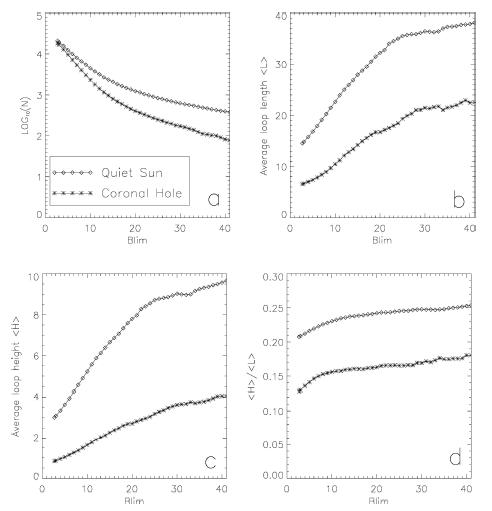}
\caption{a: The total number of loops, logarithmic scaling,
b: the average loop length
c: the average loop height
d: the ratio of average height to length vs. the minimum magnetic flux $|B_{\rm lim}|/[G]$ considered.
The rhombi correspond to quiet Sun regions and the stars to coronal hole regions in all panels.}
\label{fig8}
\end{figure}

In table \ref{table2} we present some magnetic properties of
the analysed coronal holes and quiet Sun regions.
In coronal holes the average  field strength of the
net magnetic flux $\langle B_z  \rangle$
is usually higher than in the quiet Sun and the ratio of
net unsigned magnetic field strength $\frac{\langle B_z \rangle}{\langle |Bz| \rangle}$ is higher
in coronal holes as well. $\langle B_z  \rangle$ for CHs is above $2.5$ and
$\frac{\langle B_z \rangle}{\langle |Bz| \rangle} > 0.5$, while for the QS
regions these quantities are below $1.2$ and $0.3$, respectively.

An exception is CH1, which shows untypical
features for all quantities compared with the other coronal hole regions,
and appears to have properties more similar to quiet Sun regions.
Another is the quiet Sun region $QS3$ which shows typical
coronal hole features. We expect that these regions have been misidentified
and do not consider CH1 and QS3 for the statistical investigations in the following.
Without these two regions the average values of \\
\begin{tabular}{rrl}
$\langle B_z  \rangle$ and & $\frac{\langle B_z \rangle}{\langle |Bz| \rangle}$
& are \\
$7.6{^+_-}4.8$ and & $0.77{^+_-}0.14$ & in CHs and \\
$0.4{^+_-}0.4$ and & $0.09{^+_-}0.09$ & in the QS, respectively.
\end{tabular} \\
The unbalanced flux in coronal holes (on average $77 \%$) leads
 to open field lines and to the absence
of long closed field line structures as we shall see below.

The most striking
feature of table \ref{table2} is that the average height and length
of loops is significantly smaller
in coronal holes compared with the quiet Sun. To compute average values of
loop length and height for all quiet Sun and coronal hole regions there are
two possibilities.
We compute a large number of loops from an equidistant grid of foot-points
and give each loop in different regions the same
weight. This leads to an average loop length and height of \\
\begin{tabular}{rr}
$\langle H_{CH} \rangle =0.84 Mm$, &  $\; \langle L_{CH} \rangle =6.56 Mm$ in
CHs ($18184$ loops) \\
and $\langle H_{QS} \rangle =3.02 Mm$, & $\; \langle L_{QS} \rangle =14.53 Mm$ in
QS ($21034$ loops).
\end{tabular} \\
In these statistics large regions (with many loops)
obtain  a higher weight than small regions (e.g. CH7 contains about
half of all coronal hole loops). Another possibility is to give each region the
same statistical weight and average over $\langle H \rangle $ and $\langle L \rangle $
from table \ref{table2}.
This leads to slightly different values: \\
\begin{tabular}{rr}
$\langle H_{CH} \rangle =0.87{^+_-}0.4 Mm$,&  $\; \langle L_{CH} \rangle =7.8 {^+_-}3.9 Mm$
in CHs\\
and $\langle H_{QS} \rangle =3.04{^+_-}0.7 Mm$,& $\; \langle L_{QS} \rangle =14.1{^+_-}4.1 Mm$ in
QS.
\end{tabular}

We calculated the standard deviation with $11$ (CH2-CH12)
coronal holes and $7$ (QS1-QS8, excluding QS3) quiet Sun regions.

The average loop height and length is significantly smaller
in coronal holes compared with the quiet Sun. Quiet Sun loops are on average
$3.3$ times higher and $1.8$ times longer than coronal hole loops, which
means that the average coronal hole loop is not only smaller in size than a
quiet Sun loop but also flatter in shape.
These averages have been obtained from
all loops, independently of the magnetic flux
or field strength at their foot points. As the lower limit of magnetic
field per pixel we used the effective $1- \sigma$ noise level of $2.7 G$
valid for 56-min integrated MDI magnetograms. Since it is not clear whether
the residual noise present above the $1- \sigma$ level affects the above
results, we investigated how these
values change if we consider only closed loops above a certain limit
$B_{\rm lim}$ on the statistics. This rules out loops with low magnetic flux
$|B| < B_{\rm lim}$. The results are shown in Fig. \ref{fig8}.
The number of loops (shown with logarithmic scaling)
decreases, as expected, but the decrease is much more significant for coronal hole loops
(stars) than for quiet Sun loops (rhombi). This suggest a lack of high flux closed
loops in coronal holes and is consistent with the fact that most of the coronal hole
flux, in particular in the stronger network elements,
is stored in open magnetic fields.
(This point will be analysed in greater detail in a separate paper.)
The average loop length and height increase
with  increasing  $B_{\rm lim}$, which means that, on average,
short and low closed loops have at least one footpoint in regions of low
magnetic flux.

The result that closed loops are lower and smaller in coronal holes remains
valid independently of whether we consider only loops which contain large magnetic flux
or all loops. In fact, the difference
in length and height between loops in coronal holes and quiet
Sun regions increases with $B_{\rm lim}$.
The result that coronal hole loops are on average flatter
than quiet Sun loops also remains valid independently of $B_{\rm lim}$.
The quotient $\frac{H}{L}$ increases slightly both in coronal
holes and the quiet Sun. This implies that loops starting
from weak-field regions  are a bit flatter.
This test shows that the results obtained so far are robust as
far as noise in the magnetograms is concerned.
\section{Distribution functions for loop height and length}
\label{sec4}
\begin{figure}
\hspace*{\fill}
\includegraphics[width=14cm]{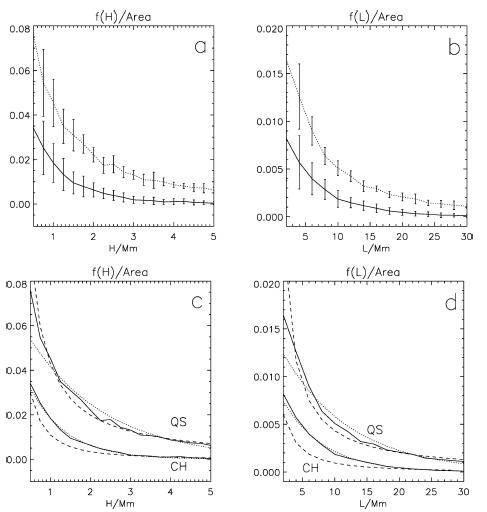}
\caption{
a:  distribution function $f(H)$ of loop height.
b: distribution function $f(L)$ of loop length.
The solid lines correspond to coronal holes and the dotted lines
to quiet Sun regions. The error bars have been calculated from the individual
statistics of the different coronal hole and quiet Sun regions.
c: distribution function $f(H)$ of loop height (solid lines), an exponential fit
(dotted lines) and a power law fit (dashed lines) are superimposed on each
observed distribution.
d: distribution function $f(L)$ of loop length (solid lines), with
exponential (dotted lines) and  power law fits (dashed lines) superimposed.}
\label{fig4}
\end{figure}
In the previous section we investigated the average height and length
of closed coronal loops in the quiet Sun and in coronal holes. Here we
calculate the distribution functions for the length and height of coronal
loops (18184 loops in 11 coronal holes and 21034 loops in 7 quiet Sun
regions). We normalize each distribution function by the area
of the investigated region (fourth column in Table \ref{table2}).
Fig. \ref{fig4}a shows the height distribution and \ref{fig4}b
the length distribution. The solid lines correspond
to coronal holes and the dotted lines to quiet Sun regions.
We first computed $H \rightarrow \frac{f(H)}{Area}$
and $L \rightarrow \frac{f(L)}{Area}$ for each
coronal hole and quiet Sun region separately. The error bars
represent the standard deviation over the 11 coronal hole and 7 quiet Sun
regions for each interval in $H$ and $L$ respectively.
The number of loops with length below a pixel diameter have not been
plotted.
The curves show firstly that the number of loops decreases rapidly with
increasing height or length of the loops and
that there are more loops (per unit area)
in quiet Sun regions compared to coronal holes,  irrespective of loop length and height.

We fit the average $f(H)$ and $f(L)$ with help of an exponential fit
($f(x)=c \cdot \exp(d \cdot x)$) and a power law fit
($f(x)=a \cdot x^b$). The best obtained fits are
(see also Fig. \ref{fig4} c and d where exponential fits are represented by
dotted lines and power law fits by dashed lines):
\begin{eqnarray*}
f_{CH}(H) =&0.055 \cdot \exp(-1.09 \cdot H) & \;\;  \\
f_{QS}(H) =&0.043 \cdot H^{-1.12} & \;\;  \\
f_{CH}(L) =&0.010 \cdot \exp(-0.16 \cdot L)& \;\;  \\
f_{QS}(L) =&0.015 \cdot \exp(-0.09 \cdot L) & \;\; \\
f_{QS}(L) =&0.052 \cdot L^{-1.08} & \;\;
\end{eqnarray*}
The exponential function fits
the coronal hole distribution functions $f_{CH}(H)$ and $f_{CH}(L)$ very well
and is superior to a power law fit, which has also been plotted for comparison.
The quiet Sun
distribution for the height $f_{QS}(H)$ is closer to a
power law distribution, while $f_{QS}(L)$ is not fit well (towards shorter
loops) by either function, even if we neglect the shortest loops, whose
number is probably influenced by the limited spatial resolution of MDI.
We have therefore given the parameters of both fits.

\begin{figure}
\hspace*{\fill}
\includegraphics[width=14cm]{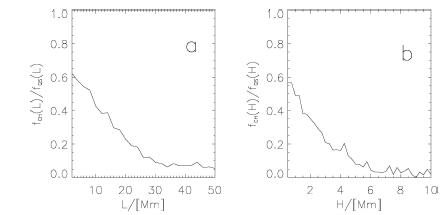}
\caption{Relative Distribution of loop length ($f_{CH}(L)/f_{QS}(L)$, left panel)
and loop height ($f_{CH}(H)/f_{QS}(H)$, right panel). The individual distribution functions
have been normalized by the area covered by the solar feature.
The  distribution functions have been computed
using all loops in quiet Sun and coronal hole regions (except CH1 and QS3).}
\label{fig7}
\end{figure}

With increasing height and length of the loops the
number of loops decreases, but remains significantly higher in the quiet Sun
than in coronal holes. It
is instructive to plot the ratio of the distributions in coronal holes and
the quiet Sun, $\frac{f_{CH}(L)}{f_{QS}(L)}$  and
$\frac{f_{CH}(H)}{f_{QS}(H)}$ (see Figs \ref{fig7} a and b, respectively).
\begin{figure}
\hspace*{\fill}
\includegraphics[width=14cm]{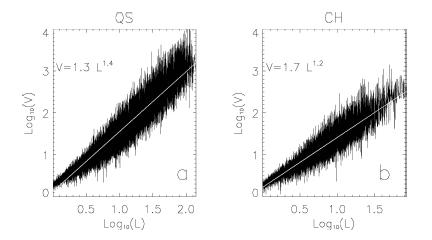}
\caption{Relation between loop length $L$ and volume $V$
(double logarithmic scaling). The solid gray lines correspond to power
law fits. The larger power law exponent obtained for QS (panel a) suggests
that these loops expand more than the CH loops of equal length (panel b).}
\label{fig_lv}
\end{figure}
The number of low and short loops in coronal holes is on average
about  $60 \%$ the number of quiet Sun loops. Towards high and long loops this ratio
drops rapidly. The number of loops longer than $30 Mm$ in coronal holes
is only about $5-10 \%$ of the number in the quiet Sun. Similarly, the number of
loops with a height of around
$4 Mm$ in coronal holes is only $10 \%$ of that in the quiet
Sun and the ratio drops further to below $5 \%$ for higher loops.

Note that although the number of longer loops drops dramatically in CHs, the
volume filled by these loops drops more slowly. We
estimate the volume $V$ filled by loops of length $L$ by
$V=\int_0^L \frac{B_{\rm Fp1}}{B(l)} dl \cdot A_{\rm Fp1}$, where
$A_{\rm Fp1}$ is the loop cross-section  area at one of the foot points,
$B_{\rm Fp1}$ the corresponding field strength
and $l$ is the distance measured along the loop.
The factor $\frac{B_{\rm Fp1}}{B(l)}$ takes into account the expansion
of the loop due to magnetic flux conservation. The relation between
loop length and volume is shown in figure \ref{fig_lv}.
The solid gray lines are power fits to the determined values.
The larger exponent obtained for the quiet Sun suggests that these
loops expand more than the CH loops of equal length. This is not surprising
given that the CH loops are flatter, i.e. do not reach so high into the
atmosphere as equivalent QS loops.
\section{Relation of loop length to temperatures.}
\label{sec5}
\begin{figure}
\hspace*{\fill}
\includegraphics[width=14cm]{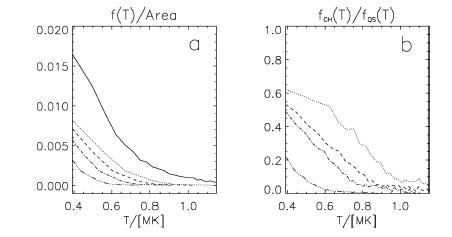}
\caption{a: Distribution function $T \rightarrow f(T)$ for QS (solid line)
and CH, were T is the temperature.
We computed the temperature from the loop length with help of the RTV-model
$(T \propto (p L)^{1/3}$. The pressure $p$ was chosen $0.11/[dyn \; cm^{-2}]$ on the quiet Sun
and $0.11/[dyn \; cm^{-2}]$ (dotted line), $0.08/[dyn \; cm^{-2}]$ (dashed line),
$0.055/[dyn \; cm^{-2}]$ (dash-dotted line), $0.03/[dyn \; cm^{-2}]$ (dash-double-dotted line)
for coronal holes. (See text).
b:  Ratio of the distribution  of loops as a function of temperature ($f_{CH}(T)/f_{QS}(T)$).
 The dotted, dash-dotted and
dash-double-dotted lines where calculated with the same  values of $p$ in the
CH as in a).}
\label{figT}
\end{figure}
\begin{figure}
\includegraphics[width=14cm]{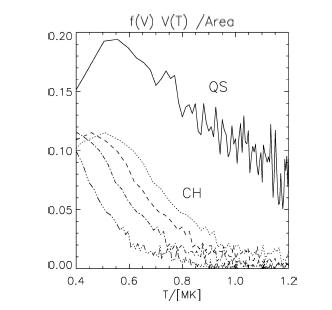}
\caption{The plot shows $f(V) \cdot V(T)$
normalized to the area of the considered region  vs. $T$.
The pressure $p$ was chosen $0.11/[dyn \; cm^{-2}]$ on the quiet Sun
and $0.11/[dyn \; cm^{-2}]$ (dotted line), $0.08/[dyn \; cm^{-2}]$ (dashed line),
$0.055/[dyn \; cm^{-2}]$ (dash-dotted line), $0.03/[dyn \; cm^{-2}]$
(dash-double-dotted line) for coronal holes.
}
\label{fig_ftv}
\end{figure}
Here we consider in a simple manner implications of the difference in
loop statistics for the temperature and temperature distribution in coronal
holes. It is not straightforward
 to draw direct quantitative conclusions regarding the emissivity,
however. While one might assume that most of the emission comes from
closed magnetic loops, it is likely that  some emission is
related to open fields. While in quiet Sun regions (where high large loops
exist) the emission from open field lines can probably be neglected,
this might not be
the case for coronal holes. In a region without (or with very few) high and
long loops the emission from the dilute plasma on open field lines might
become relatively important. The overall emissivity will, of course, still
be very small.

In the following we use the assumption
that short loops are cooler than long loops. This assumption is supported
by a hydrostatic model \cite{rosner78} (RTV-model) which provides the scaling law
\begin{equation}
T_{max}=1400 \cdot (pL)^{1/3},
\label{rtv}
\end{equation}
where $T_{max}$ is the maximum temperature, $L$ the loop length in $cm$
and $p$ the pressure in $dyn \; cm^{-2}$. We use this scaling law to
approximate the temperatures of the loops calculated here.
We assume that the temperature along the loops equals $T_{\rm max}$.
This agrees with the thermal structure of hot loops deduced from CDS
(e.g. \inlinecite{brkovic02}).
\inlinecite{maxson77} found a pressure of $0.11 \; dyn \; cm^{-2}$ for large-scale
structures and $0.03 \dots 0.08 \; dyn \; cm^{-2}$ in coronal holes.
Using these scaling laws and $\langle L_{CH} \rangle =7.8 \; Mm $ and
$\langle L_{QS} \rangle =14.1 \; Mm $
we find $\frac{T_{QS}}{T_{CH}}=1.2$ if the pressure is assumed to be the same
in coronal holes and the quiet Sun. If the pressure is kept fixed to $0.11$
in the quiet Sun and reduced to $0.08,\; 0.055,\; 0.03$ (maximum, average and
minimum value in coronal holes) we find an increase of the
average temperature ratio to $\frac{T_{QS}}{T_{CH}}=1.4, \; 1.5 \;, 1.9,$
respectively. Since the \inlinecite{maxson77} values for the CHs refer
to high temperatures and probably mix contributions of both open and
closed field lines, we suspect that the most realistic result is obtained
assuming the pressure in CH loops to be the same as in QS loops. We have
neglected any dependence of $p$ on $T$ or $L$.

The effect on the average temperature ratio may not look very dramatic,
but the absence of long loops in coronal holes has a dramatic effect on
the temperature distribution function, as can be seen from
Fig. \ref{figT}. The number of loops decreases rapidly with temperature
for both QS and CH regions, irrespective of the assumed pressure in
CH loops. Since longer loops are hotter, the total emitting volume does
not as rapidly decrease as suggested by Fig. \ref{figT} a).

We find that hot loops are
practically absent in coronal holes . The effect is clearly visible already if the
same scaling law (same value of $p$)is applied to coronal holes and
the quiet Sun. It becomes even stronger when
we reduce $p$ in coronal holes following
\inlinecite{maxson77}.
Figure \ref{figT} (panel b) shows the decrease in the ratio of loops (per area)
$\frac{f_{CH}}{f_{QS}}$ with increasing loop temperature.
 We find that (depending on the assumed $p$)
$\frac{f_{CH}}{f_{QS}}$ has a ratio of $0.62, \; 0.53, \; 0.49, \; 0.20$
at a temperature of $0.4 MK$ and the ratio reduces to
$0.18, \; 0.06, \; 0.05, \; 0.00$ respectively at a temperature of $0.9 MK$.
Of greater interest for the observations than just the number of loops at a
given temperature is the emitting volume filled by gas at that temperature.
Using the fits plotted in Fig. \ref{fig_lv} we plot $f(V) \cdot V(T)$
normalized to the area of the considered region in figure
\ref{fig_ftv} vs. $T$. Again, the very large contrast at high temperatures
and the much smaller contrast at low temperatures are visible.

One has to recognize, however, that the temperatures calculated here with
scaling laws can only
be a rough estimate. The values of $p$ required for the RTV-model (\ref{rtv})
are not know very precisely. In addition, the RTV-model has
been derived assuming a static equilibrium (no plasma flow) along the
loops, zero gravity, uniform heating and a constant cross section.
These assumptions are not exact.
\citeauthor{serio81} \citeyear{serio81} extended the RTV-model, taking into
account gravity and non uniform heating. The scaling laws are similar to
the RTV-model  ($T=1400 \cdot (pL)^{1/3} \cdot \exp \left[-0.04L \;
\left(2/s_H+1/s_p \right) \right]$ with the heating deposition scale height $s_H$ and
the pressure scale height $s_p$.)
The gravity can probably be
neglected here as we deal with loops small compared with the pressure
scale height.
\inlinecite{kano95} showed that the
heating deposition distribution along the loop influences the scaling laws.
By assuming that the heat input occurs at the top of the loop
\citeauthor{kano95} \citeyear{kano95}
also got the RTV-scaling law (\ref{rtv}), but if the heat input is constant
along the loop the authors found $T_{\mbox{max}}=1100 \cdot (pL)^{1/3}$. If we
would use the latter model instead of RTV the loop temperatures would be reduced
by a factor of $0.8$. The temperature ratios and the shape of the
distribution functions would remain the same, however. Another modification
of the RTV-model, e.g. a different loss function \cite{kano95} lead to
$T_{\mbox{max}} \propto (pL)^{\gamma}$ with $\gamma$ different from the
RTV-value of $1/3$, but these modifications are related to temperatures
above $10^6 K$ and it is not clear to which extent they apply here.
\inlinecite{aschwanden00} found analytic
approximations for hydrostatic loops by fitting results of a numerical code.
The authors mainly concentrated on hot ($1-10MK$) loops and the temperature
distribution along the loops, which is outside the scope of this paper.
\inlinecite{winebarger02} have argued that loops are inherently dynamic.
Again, such an extension is well beyond the scope of the present paper.
\section{What magnetic features are responsible for coronal holes?}
\label{sec6}
\begin{figure}
\includegraphics[width=9cm]{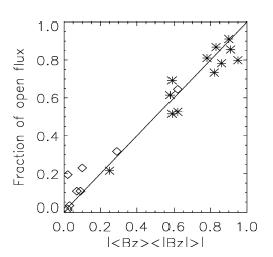}
\caption{Fraction of open flux, normalized to the total flux of
the studied region vs.
$\frac{\langle B_z \rangle}{\langle |B_z| \rangle}$.
The correlation coefficient between these
quantities is $0.98$. The rhombi correspond to QS and the stars to CH.
The solid line corresponds to
$\frac{\langle B_z \rangle}{\langle |B_z| \rangle}=$ fraction of open flux.
}
\label{fraction_open}
\end{figure}
\begin{figure}
\hspace*{\fill}
\includegraphics[width=14cm]{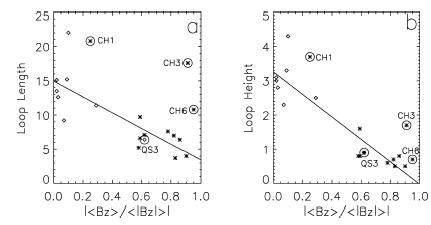}
\caption{The figure shows the relative net flux $\frac{\langle B_z  \rangle}{\langle |B_z| \rangle}$
against the average loop length (left panel) and average loop height
(right panel). Each symbol represents one region.
The rhombi correspond to the quiet Sun regions and the stars to
coronal holes. The lines are a linear fit (CH1, CH3, CH6 and QS3 are shown in
the figures (and marked with circles), but have not been considered for the linear fit.).}
\label{fig6}
\end{figure}
From previous publications \cite{harvey82,deforest97,wilhelm00,belenko01}
we know that the magnetic flux in coronal holes is dominated by one polarity and
consequently the magnetic field in a coronal hole is not flux
balanced. This result is confirmed by  Table \ref{table2}.
On average $\frac{\langle B_z \rangle}{\langle |B_z| \rangle}=0.77$
for the studied CHs and $\frac{\langle B_z \rangle}{\langle |B_z| \rangle}=0.09 $
for the QS regions. Except for one region each, all CHs are distinctly
different from the QS regions.

Here we investigate how  $\frac{\langle B_z  \rangle}{\langle |B_z| \rangle}$
is related to the amount of open flux,
the average loop length $\langle L \rangle $ and height $\langle H \rangle $.

Figure \ref{fraction_open} shows the amount of open flux, normalized to
the total flux of the studied region.
The rhombi correspond
to quiet Sun regions and the stars to coronal holes. (Every rhombus and
every star represents one quiet Sun and coronal hole region, respectively.)
Remarkably, there is a small scatter (correlation coefficient $0.98$).
The reason for the scatter is, that a part of the
unbalanced flux in a particular region is closed and connects to fields
outside the area entering the statistics (e.g. loops crossing the
boundary of the CH.)

Figures \ref{fig6} a) and b) show the average loop length and
the average loop height, respectively, vs. the relative flux balance.

There is a clear trend for the loop height to decrease with
$\frac{\langle B_z  \rangle}{\langle |B_z| \rangle}$ ,
the correlation coefficient between these two quantities being $r_c=-0.87$.
Interestingly the loop length is not so strongly correlated
($r_c=-0.57$), mainly because of CH3 and CH6, which have anomalously long
loops (but not strikingly high ones).
The strongest correlation is found between
$\frac{\langle B_z  \rangle}{\langle |B_z| \rangle}$ and
$\frac{\langle H  \rangle}{\langle L  \rangle}$, the correlation coefficient
being $-0.92$. We find that the average unsigned magnetic field strength is
about a factor of two higher in coronal holes for the regions investigated
here ($\langle|B_z| \rangle=9.3 {^+_-} 4.6 G$ in CHs and
$\langle|B_z| \rangle=4.5 {^+_-} 0.9 G$ in QS).
\section{Conlusions}
\label{sec7}
We have investigated the local magnetic field structure of
equatorial coronal holes and its relation to the emissivity.
Coronal holes have a higher relative background magnetic
field than quiet Sun regions and consequently most of
the total magnetic flux (usually some $80 \%$, for some
holes more than $90 \%$ and for others only about $60 \%$)
is unbalanced flux which mainly results
in open magnetic field lines. (Some of it may be balanced
in nearby regions lying outside the boundary of the CH.)

Closed loops exist, however,
also in coronal holes but their average
length and height  is lower than in the quiet Sun.
These results might provide an understanding of the appearance of coronal holes
at different wavelength.
The number of low and short loops in coronal holes is only somewhat smaller
than in the quiet Sun (the ratio being about $65 \%$),
but for long and high loops the
ratio drops to less than  $10 \%$.
We also find that the ratio of loop height to loop length is lower in
coronal holes, so that coronal hole loops are flatter.
Converting loop length into temperature using the Rosner-Tucker-Vaiana
scaling law we obtain that the fraction of the volume filled with transition
region gas (at $T=4 \, 10^5 K$) in CHs is on average $70 \%$ of that in the QS,
while at temperatures of $0.9 MK$ this fraction drops to $10 \%$
(see Fig. \ref{fig_ftv}).

The local magnetic field structure thus provides  an explanation
why hot coronal line emission is absent in coronal holes. Due to
the almost complete absence of closed
loops at coronal temperatures in coronal holes there is hardly any
emission in lines formed at high temperatures.
The current, simple analysis does not provide a full solution
of the puzzle yet as why CHs are just as bright as QS at transition
region temperatures. Although the volume filling due to cool loops is a large
fraction of that in the quiet Sun, it is still significantly lower. This
may partly be due to the limited resolution and sensitivity of MDI, so
that we underestimate the number of short loops. However, there would still
appear to be some need for significant chromospheric and transition region
heating along the open field lines in order to reproduce the observations.
We must also keep in mind that on average the field is a bit stronger in CHs,
so that the open flux need not be quite as bright as the closed, even in TR.

For future work it would be interesting to investigate also
the local magnetic field structure of polar coronal
holes. Unfortunately, magnetic field measurements of the
polar magnetic field are not available to the necessary precision yet.
The Solar Orbiter mission \cite{marsch01} will provide high quality
measurements of polar magnetic fields.

\acknowledgements
We thank Bernd Inhester for useful discussions and Natalie Krivova
for providing us with the 56-min averaged MDI-magnetograms.
We used coronal hole maps prepared at NSO/Kitt Peak by Karen Harvey
and Frank Recely as part of an NSF grant. We thank Andreas Kopp for his
assistance in converting these PDF-file maps into vector graphics.
This work was supported by  DLR-grant 50 OC 0007.
We thank the referee Leon Golub for useful remarks.
\clearpage
\appendix
\renewcommand{\theequation}{\Alph{section}.\arabic{equation}}
\section{The magnetic field model}
\label{appendixb}
 \begin{table}
\caption{List of coronal hole and quiet Sun regions. The first column stands for the
coronal hole CH1-12 and quiet Sun QS1-8 regions investigated. The second and third
columns contain the MDI number of the full disk magnetogram and observation date
respectively. We chose  the full disk magnetograms in order to have the coronal holes
closest to the central meridian. The last four columns contain the positions of the
investigated region in MDI-pixels.}
\begin{tabular}{|r|r|r|r|r|r|r|}
\hline
Region & MDI-N. & Date & xmin & xmax & ymin & ymax \\
\hline
CH1 & 73058 & 0503 & $     406$ & $     528$ & $     478$ & $     572$ \\
CH2 & 73082 & 0504 & $     358$ & $     613$ & $     259$ & $     462$ \\
CH3 & 73010 & 0501 & $     406$ & $     526$ & $     457$ & $     644$ \\
CH4 & 72962 & 0429 & $     425$ & $     576$ & $     660$ & $     815$ \\
CH5 & 72866 & 0425 & $     417$ & $     632$ & $     767$ & $     906$ \\
CH6 & 72818 & 0423 & $     417$ & $     512$ & $     369$ & $     486$ \\
CH7 & 72842 & 0424 & $     260$ & $     746$ & $      96$ & $     385$ \\
CH8 & 72746 & 0420 & $     430$ & $     560$ & $     235$ & $     334$ \\
CH9 & 72722 & 0419 & $     467$ & $     589$ & $     668$ & $     759$ \\
CH10 & 72650 & 0416 & $     491$ & $     589$ & $     604$ & $     695$ \\
CH11 & 72650 & 0416 & $     398$ & $     542$ & $     264$ & $     465$ \\
CH12 & 72506 & 0410 & $     462$ & $     547$ & $     676$ & $     778$ \\
\hline
QS1 & 73058 & 0503 & $     500$ & $     699$ & $     250$ & $     429$ \\
QS2 & 73010 & 0501 & $     500$ & $     629$ & $     400$ & $     529$ \\
QS3 & 72962 & 0429 & $     400$ & $     599$ & $     220$ & $     379$ \\
QS4 & 72818 & 0423 & $     450$ & $     549$ & $     500$ & $     599$ \\
QS5 & 72770 & 0421 & $     580$ & $     679$ & $     720$ & $     819$ \\
QS6 & 72650 & 0416 & $     600$ & $     699$ & $     700$ & $     799$ \\
QS7 & 72842 & 0424 & $     470$ & $     609$ & $      70$ & $     169$ \\
QS8 & 72842 & 0424 & $     350$ & $     449$ & $     286$ & $     385$ \\
\hline
\end{tabular}
\label{tableA}
\end{table}
We use the line of sight magnetic field observed with SOHO/MDI
and reconstruct the coronal magnetic field as a potential field
with the help of a Greens function method \cite{aly89}.
The potential and magnetic field are presented as
\begin{eqnarray}
\Phi({\bf r}) & = &-\frac{1}{2 \pi} \int_{\partial \Omega} B_z({\bf r^{'}})
\, \frac{d \sigma^{'}}{|{\bf r}-{\bf r^{'}}|}, \\
{\bf B} & = & \nabla \Phi,
\end{eqnarray}
where $\Phi$ is the scalar potential, $B_z$ the line of sight photospheric
magnetic field, $\partial \Omega$ the bottom boundary (photosphere) of the computational
box, $d \sigma = dx \, dy$, ${\bf r}=\sqrt{x^2+y^2+z^2}$ and ${\bf B}$ is
the 3D magnetic field.

We use MDI-data as input for our potential magnetic field code.
 The noise-level of the original MDI-data
has been reduced by forming from
56 consecutive 1-min magnetograms, from which 56 min averages are constructed, after
compensating for solar (differential) rotation (see \inlinecite{krivova04}).
To avoid the influence of the remaining noise in the
magnetic field data, values of $|Bz| < 2.7G $ are set to zero, as well
as isolated data points (data points with $|Bz| \ge 2.7 G$ but with no
neighbour points with the same polarity). The resolution of one MDI-pixel
is approximately $1.4$Mm. We compute the potential magnetic field in
rectangular boxes.
The magnetic field lines are
calculated with a fourth order Runge-Kutta field line tracer with step size
control. The field line tracer interpolates the magnetic field between
pixels with a trilinear interpolation routine.
The pixel size in our computational box corresponds to one MDI pixel.
We start the field
line integration on the photosphere and integrated the magnetic field line
until it reaches again the photosphere (closed field lines, loops) or the
upper boundary of the computational box. These field lines are interpreted
as open field lines. This interpretation seems to be quite well justified
as the fraction of open flux computed with help of the so defined open field
lines and the unbalanced magnetic flux on the photosphere coincide quite well
(see figure \ref{fraction_open}.

  To diminish the influence of lateral boundary conditions
we do not start the field line integration in a lateral boundary layer of 20
MDI pixels.

\end{article}
\end{document}